\documentclass[english]{article}
\usepackage[T1]{fontenc}
\usepackage[latin9]{inputenc}
\usepackage{xcolor}
\usepackage{pdfcolmk}
\usepackage{booktabs}
\usepackage{units}
\usepackage{url}
\usepackage{amsmath}
\usepackage{amssymb}
\usepackage{graphicx}  
\usepackage{a4wide}
\usepackage{geometry}
\makeatletter

\newcommand{\lyxmathsym}[1]{\ifmmode\begingroup\def\b@ld{bold}
	\text{\ifx\math@version\b@ld\bfseries\fi#1}\endgroup\else#1\fi}

\providecommand{\tabularnewline}{\\}
\providecolor{lyxadded}{rgb}{0,0,1}
\providecolor{lyxdeleted}{rgb}{1,0,0}

\DeclareRobustCommand{\lyxsout}[1]{\ifx\\#1\else\sout{#1}\fi}

\numberwithin{equation}{section}
\numberwithin{figure}{section}

\usepackage{placeins}
\usepackage{hyperref}
\usepackage{babel} 
\usepackage{hyphenat} 

\@ifundefined{showcaptionsetup}{}{%
	\PassOptionsToPackage{caption=false}{subfig}}
\usepackage{subfig}
\makeatother

\usepackage{babel}
\begin{document}
	
	\title{Gaussian Process Regression for Pricing Variable Annuities with Stochastic
		Volatility and Interest Rate}
	
	
	\author{ \textsc{Ludovic Gouden\`ege}\thanks{F\'ederation de Math\'ematiques de CentraleSup\'elec - CNRS FR3487, France -\texttt{ ludovic.goudenege@math.cnrs.fr}} \and \textsc{Andrea Molent}\thanks{Dipartimento di Scienze Economiche e Statistiche, Universit\`a degli Studi di Udine, Italy - \texttt{andrea.molent@uniud.it}} \and \textsc{Antonino Zanette}\thanks{Dipartimento di Scienze Economiche e Statistiche, Universit\`a degli Studi di Udine, Italy - \texttt{antonino.zanette@uniud.it}}}
	\date{}
	 
	\maketitle
	\noindent\rule{1\columnwidth}{1pt}
	\begin{abstract}
	 In this paper we investigate price and Greeks computation of a Guaranteed Minimum Withdrawal Benefit (GMWB) Variable Annuity (VA) when both stochastic volatility and stochastic interest rate are considered together in the Heston Hull-White model. We consider a numerical method the solves the dynamic control problem due to the computing of the optimal withdrawal. Moreover, in order to speed up the computation, we employ Gaussian Process Regression (GPR). Starting from observed prices previously computed for some known combinations of model parameters, it is possible to approximate the whole price function on a defined domain. The regression algorithm consists of algorithm training and evaluation. The first step is the most time demanding, but it needs to be performed only once, while the latter is very fast and it requires to be performed only when predicting the target function. The developed method, as well as for the calculation of prices and Greeks, can also be employed to compute the no-arbitrage fee, which is a common practice in the Variable Annuities sector. Numerical experiments show that the accuracy of the values estimated by GPR is high with very low computational cost. Finally, we stress out that the analysis is carried out for a GMWB annuity but it could be generalized to other insurance products. 
	\end{abstract}
	
	\noindent \emph{\large{}Keywords}:
		GMWB pricing, Heston Hull-White model, Numerical method, Machine Learning, Gaussian Process Regression
	  
	\noindent\rule{1\columnwidth}{1pt}
	\section{Introduction}
	
	In this paper we focus on a particular Variable Annuity (VA) which
	leads the living benefit riders: the Guaranteed Minimum Withdrawal
	Benefit (GMWB). This contract is initiated by making a lump sum payment
	which is then invested in risky assets, usually a mutual fund. The
	holder of the policy (PH) is entitled to withdraw a fixed amount at
	the contract anniversaries, even if the risky account has declined
	to zero. If the amount withdrawn exceeds the guaranteed amount then
	a penalty is imposed. The contract also includes the payment of a
	benefit to the heirs of the PH in case of death during the coverage
	period. When the contract reaches maturity, a final payoff is paid
	out and the contract ends.
	
	Researchers have engaged to develop new numerical techniques that
	are increasingly efficient and less computationally expensive. For
	example, Bacinello et al. \cite{bacinello2011} consider Monte Carlo
	methods to price and compute the Greeks of GMWB products. Chen and
	Forsyth \cite{chen2008} and Donnelly et al. \cite{donnelly2012}
	introduce partial derivative equations, while Costabile \cite{costabile2017}
	propose pricing methods based on the use of trees. More recently,
	Gouden\`ege et al. apply the Hybrid Tree-PDE method in order to compute
	the price and the Greeks of GLWB and GMWB contracts \cite{goudenege2016,goudenege2019}.
	However, as these techniques are more and more efficient, calculation
	times remain non-negligible, especially when the stochastic model
	considered to represent the market dynamics includes many random factors,
	such as stochastic volatility and stochastic interest rate. Moreover,
	the use of an accurate model that includes stochastic volatility and
	stochastic interest rate is crucial for treating products with long
	maturity, such as VAs in general.
	
	In this paper, we investigate the evaluation of a GMWB contract when
	the Heston Hull-White model is considered. Such a stochastic model
	is particularly interesting since it provides for stochastic volatility
	and interest rate together. In particular, we consider the Hybrid
	Tree-PDE (HPDE) approach introduced by Briani et al. \cite{briani2018}
	which is a backward induction algorithm that works following a finite
	difference PDE method in the direction of the share process and a
	tree method in the direction of the other random sources, i.e. the
	stochastic volatility and the stochastic interest rate. Moreover,
	we develop a new trinomial tree to diffuse the additional random sources
	and we employ it to obtain an improved version of the HPDE method.
	Such a numerical method turns out to be particularly suitable to the
	scope of GMWB products since it is able to treat the long maturity
	of these insurance products. Moreover, since the PH can select the
	optimal amount to withdraw at each contract anniversary, a dynamic
	control problem must be solved and the HPDE method can be effectively
	applied to deal with this problem. In particular, we proceed backward
	in time by employing the HPDE method to compute the continuation value
	of the contract between two contract anniversaries and by modifying
	the contract state variables at each anniversary according to the
	optimal withdrawal strategy. To the best of our knowledge this is
	the first time that the GMWB has been priced in the Heston Hull-White
	model also considering an optimal withdrawal strategy.
	
	The computations that insurance companies face depend on model parameters
	and contract parameters which are only few and vary in a small range.
	This context is similar to what happens in the derivatives world where
	there are many repeated valuations of standardized products. An innovative
	solution to speed up derivative pricing has been proposed by De Spiegeleer
	et al. \cite{deSpiegeleer2018}, which consists in employing Machine
	Learning techniques to predict the price of the derivatives from a
	training set made of observed prices for particular combinations of
	model parameters. In particular, they consider Gaussian Process Regression
	(GPR) which is a Bayesian non-parametric technique. All the information
	needed to compute a function (the price or the Greeks of a derivative)
	can be summarized in a training set, then the algorithm learns the
	function and it can be applied to make predictions for new input parameters
	very fast. A similar approach, together with clustering techniques,
	has been employed by Gan \cite{gan2013} to price large portfolios
	of VAs in the Black-Scholes model. Unlike what De Spiegeleer et al.
	do, Gan considers fixed values for the market parameters (interest
	rate and volatility), having to repeat the procedure at each change
	in market conditions. Gan and Lin \cite{gan2015} propose a novel
	approach that combines clustering technique and GPR to efficiently
	evaluate policies considering nested simulations. More recently, Gan
	and Lin \cite{gan2017} propose a two-level metamodelling approach
	to efficiently estimating the sensitivities required to hedge a portfolio
	of VA, by employing GPR and Monte Carlo simulations. 
	
	As a second outcome, we show that Machine Learning techniques can
	be applied in the scope of GMWB products as far as pricing and Greeks
	calculation are considered in the Heston Hull-White model. In particular,
	we show that it is possible to develop a regression method capable
	of predicting the prices and the Greeks of each policy, for each combination
	of market and contract parameters within the training range. The training
	procedure requires the accurate calculation of the price and of the
	Greeks of a policy for different parameters configurations, which
	is done by the HPDE method. The training procedure needs to be performed
	only once and the developed regression can be applied under various
	market conditions and for different GMWB contracts. Moreover we also
	show that GPR can be applied effectively to compute the no-arbitrage
	fee, that is the cost that makes the contract fair under a risk neutral
	probability. Numerical tests show that the use of the GPR method reduces
	the computational time by several factors with a small loss of accuracy
	which makes the technique particularly interesting. Furthermore, unlike
	what proposed by Gan \cite{gan2013}, no clustering technique for
	portfolios is required, providing and universal method that does not
	depends on specific template policies.
	
	The main results of this paper consist in applying the HPDE method
	to appraising GMWB products and in speeding-up the evaluation by means
	of the GPR method so to obtain accurate evaluation of the policies
	in a very short time, what is required by all insurance companies.
	The reminder of the paper is organized as follows. In Section 2 we
	present the GMWB contract and we explain how to employ HPDE to compute
	prices and Greeks. In Section 3 we briefly review the GPR method and
	we explain how to apply it to the GMWB scope. In Section 4 we report
	the results of some numerical tests. Finally, Section 5 draws the
	conclusions.
	
	\section{The GMWB Contract and its Appraisal}
	
	In this Section we present the GMWB contract and the numerical method.
	Following Chen and Forsyth \cite{chen2008} and Gouden\`ege et al. \cite{goudenege2019},
	we focus on a simplified version of the contract.
	
	\subsection{The GMWB Contract}
	
	At contract inception the PH pays, as a lump sum, the premium $P$
	to the insurance company. Then, the PH is entitled to perform withdrawals
	at each contract anniversary, starting from $t=1$. In case of death
	of the insured, his heirs receive a benefit and the contact ends.
	If alive at maturity, the PH performs the last withdrawal, receives
	a final payoff and the contract ends. The contract evolution is described
	by two state parameters, the account value ($A_{t}$, $A_{0}=P$)
	and the base benefit ($B_{t}$, $B_{0}=P$), which change during the
	contract life according to the underlying fund value $S_{t}$ and
	to the PH withdrawals. Specifically, during the time between two consecutive
	contract anniversaries, the base benefit $B_{t}$ does not change,
	while the account value $A_{t}$ follows the same dynamics of $S_{t}$,
	with the exception that some fees, determined by the parameter $\alpha$,
	are subtracted from $A_{t}$ as follows:
	
	\begin{equation}
	dA_{t}=\frac{A_{t}}{S_{t}}dS_{t}-\alpha A_{t}dt.\label{eq:fees_cont}
	\end{equation}
	During the contract anniversaries, the PH can withdraw from his account
	a minimum amount $G$ which is usually equal to $P/T$. Let us denote
	$A_{t_{i}^{\left(-\right)}},B_{t_{i}^{\left(-\right)}}$ the contract
	state variables just before the i-th contract anniversary occurs at
	time $t_{i}=i$, and $A_{t_{i}^{\left(+\right)}},B_{t_{i}^{\left(+\right)}}$
	the same state variables just after $t_{i}$ occurs. Moreover, let
	$w_{i}$ represent the amount withdrawn at time $t_{i}$, which is
	required to be non-negative and smaller than the base benefit $B_{t_{i}^{\left(-\right)}}$.
	If the amount withdrawn satisfies $w_{i}\leq G$, then no penalty
	is imposed, whereas, if $w_{i}>G$, a proportional penalty charge
	$\kappa\left(w_{i}-G\right)$ reduces the amount actually received
	by the PH. To this aim, we denote with $f_{i}\left(w_{i}\right):\left[0,B_{t_{i}^{\left(-\right)}}\right]\rightarrow\mathbb{R}$
	the net cash flow at time $t_{i}$:
	\[
	f_{i}\left(W_{i}\right)=\begin{cases}
	w_{i} & \mbox{if }w_{i}\leq G\\
	w_{i}-\kappa\left(w_{i}-G\right) & \mbox{if }w_{i}>G.
	\end{cases}
	\]
	The contract state variables just after the withdrawal are updated
	as follows:
	\begin{equation}
	A_{t_{i}^{\left(+\right)}}=\max\left(A_{t_{i}^{\left(-\right)}}-w_{i},0\right),\text{ and }B_{t_{i}^{\left(+\right)}}=B_{t_{i}^{\left(-\right)}}-w_{i}.\label{eq:update}
	\end{equation}
	If the PH passes away at time $\tau\in\left]t_{i-1},t_{i}\right[$,
	then his heirs receive a death benefit which is worth 
	\begin{equation}
	DB=\max\left(A_{t_{i}^{\left(-\right)}},\left(1-\kappa\right)B_{t_{i}^{\left(-\right)}}\right)
	\end{equation}
	and the contract ends. If the PH is alive at maturity time $T$, he
	is entitled to perform the last withdrawal and then he receives the
	final payoff which is worth
	\begin{equation}
	FP=\max\left(A_{T^{\left(+\right)}},\left(1-\kappa\right)B_{T^{\left(+\right)}}\right),\label{eq:FP}
	\end{equation}
	and the contract terminates. The effects of the mortality on the contract
	are described by means of two function: $\mathcal{M}:\left[0,T\right]\rightarrow\mathbb{R}$,
	which stands for the probability density of the random death year
	of the PH, and $\mathcal{R}:\left[0,T\right]\rightarrow\mathbb{R}$
	that represents the fraction of the original owners who are still
	alive at time $t$, that is $\mathcal{R}\left(t\right)=1-\int_{0}^{t}\mathcal{M}\left(s\right)ds.$
	For seek of simplicity, we assume $\mathcal{M}$ to be constant between
	contract anniversaries, and thus it can be easily obtained from a
	Mortality Table.
	
	Let $\mathcal{V}\left(A,B,v,r,t_{i}^{\left(-\right)}\right)$, $\mathcal{V}\left(A,B,v,r,t_{i}^{\left(+\right)}\right)$
	represent the contract value immediately before and after performing
	the withdrawal at time $t_{i}$, respectively. Following Forsyth and
	Vetzal \cite{forsyth2014optimal}, we consider the worst withdrawal
	strategy for the insurer, that is 
	\begin{equation}
	w_{i}=\underset{w\in\left[0,B_{t_{i}^{\left(-\right)}}\right]}{\mbox{argmax}}\ \mathcal{V}\left(\max\left(A_{t_{i}^{\left(-\right)}}-w,0\right),B_{t_{i}^{\left(-\right)}}-w,v_{t_{i}},r_{t_{i}},t_{i}^{\left(+\right)}\right)+\mathcal{R}\left(t_{i}\right)f_{i}\left(w\right).\label{eq:W}
	\end{equation}
	Assuming such a strategy and hedging against it is clearly very conservative,
	but it ensures no losses for the insurance company. 
	
	\subsection{The Stochastic Model for the Underlying and the GMWB Appraisal}
	
	The Heston Hull-White model includes both stochastic volatility and
	stochastic interest rate as follows:
	
	\begin{equation}
	\begin{cases}
	dS_{t} & =r_{t}S_{t}dt+\sqrt{v_{t}}S_{t}dZ_{t}\\
	dv_{t} & =k_{v}\left(\theta_{v}-v_{t}\right)dt+\omega_{v}\sqrt{v_{t}}dW_{t}^{v},\\
	dr_{t} & =k_{r}\left(\theta_{r}\left(t\right)-r_{t}\right)dt+\omega_{r}dW_{t}^{r},
	\end{cases}\label{eq:HHW}
	\end{equation}
	where $Z$, $W^{v}$ and $W^{r}$ are Brownian motions such that $d\left\langle Z_{t},W_{t}^{v}\right\rangle =\rho_{v}dt$,
	$d\left\langle Z_{t},W_{t}^{r}\right\rangle =\rho_{r}dt$ and $d\left\langle W_{t}^{v},W_{t}^{r}\right\rangle =0$.
	Here $\theta_{r}\left(t\right)$ is a deterministic function which
	is completely determined by the market values of the zero-coupon bonds
	by calibration. For seek of simplicity, we assume that the market
	price of a zero-coupon bond at time $t$ with maturity $\bar{t}$
	is given by $P^{M}\left(t,\bar{t}\right)=e^{-r_{0}\left(\bar{t}-t\right)}$,
	the so-called \emph{flat curve} case.
	
	The parameters which identify the Heston Hull-White model are the
	initial volatility $v_{0}$, the rate of mean reversion $k_{v}$,
	the long run variance $\theta_{v}$, the volatility of volatility
	$\omega_{v}$, the correlation $\rho_{v}$, the initial interest rate
	$r_{0}$, the rate of mean reversion $k_{r}$, the volatility of the
	interest rate $\omega_{r}$ and the correlation $\rho_{r}$. In order
	to compute the fair price and the Greeks of a GMWB contract in the
	Heston Hull-White model, we apply the HPDE method, introduced by Briani
	et al. \cite{briani2018} for option pricing and successfully employed
	in the framework of VAs by Gouden\`ege et al. \cite{goudenege2016,goudenege2019}
	while considering the Heston and the Black-Scholes Hull-White models.
	The HPDE is a backward induction algorithm that works following a
	finite difference PDE method in the direction of the share process
	and following a tree method in the direction of the volatility and
	of the interest rate. We give here some details about the numerical
	method and we refer the interested reader to \cite{briani2017,briani2018,goudenege2016,goudenege2019}
	for a detailed description. We start by observing that $r$ can be
	written as
	\begin{equation}
	r_{t}=\omega_{r}x_{t}+\varphi_{t},
	\end{equation}
	where 
	\begin{equation}
	x_{t}=-k_{r}\int_{0}^{t}x_{t}ds+W_{t}^{r}\ \text{ and }\ \varphi_{t}=r_{0}e^{-k_{r}t}+k_{r}\int_{0}^{t}\theta_{r}\left(s\right)e^{-k_{r}\left(t-s\right)}ds.
	\end{equation}
	We consider now an approximating tree $\left(\hat{v}_{n}^{h}\right)_{n\in\left\{ 0,\dots,N\right\} }$
	for $v$ and $\left(\hat{x}_{n}^{h}\right)_{n\in\left\{ 0,\dots,N\right\} }$
	for $x$, both with $N$ time-steps. Since the GMWB contract are long
	maturity contracts (up to $30$ years), the binomial trees proposed
	in \cite{briani2018} are not suitable in this context. In the Appendix
	\ref{sec:Ap} we report how to obtain trinomial trees that are more
	suitable for the scope of VAs. These tree are a good compromise between
	precision and required working memory. So, let $\Delta t=\frac{T}{N}$
	and $t_{n}=n\Delta t$. We proceed backward along the tree and at
	each time $t_{n}$, for each couple of nodes $\left(\hat{v}_{n}^{h},\hat{x}_{n}^{h}\right)$,
	we consider the values of $v_{t}$ and $x_{t}$ equal to $\hat{v}_{n}^{h}$
	and $\hat{x}_{n}^{h}$ respectively. Then, we compute the contract
	value at time $t_{n}$ for each value of $B$ as the solution of a
	one-dimensional partial derivative equation that depends on a suitable
	transformation $z$ of the account value $A$ (see Briani et al \cite{briani2018}):
	\begin{equation}
	\begin{cases}
	\partial_{t}u+\mu\left(v,x,t\right)\partial_{z}u+\frac{1}{2}\rho_{3}^{2}v\partial_{z}^{2}u & =0\\
	u\left(t_{i+1},z;v,x\right) & =\mathcal{V}\left(A\left(z\right),B,v,\omega_{r}x+\varphi_{t},t_{i+1}^{-}\right)
	\end{cases}
	\end{equation}
	The solution of such a PDE is computed at each tree node by performing
	a finite difference step. 
	
	\section{Gaussian Process Regression for GMWB}
	
	In this Section, we present a brief review of the Gaussian Process
	Regression and for a comprehensive treatment we refer to Rasmussen
	and Williams \cite{Rasmussen2006}. GPR is a class of non-parametric
	kernel-based probabilistic models which represents the input data
	as the random observations of a Gaussian stochastic process. In a
	nutshell, the GPR technique is used to make extrapolations in high
	dimension. The most important advantage of this approach is that it
	is possible to effectively exploit a complex dataset which may consist
	of points sampled randomly in a multidimensional space. This is particularly
	useful when the domain dimension is high, in which case even the mere
	construction of a grid of points requires a prohibitive computational
	effort. Finally, compared to other artificial intelligence techniques
	such as neural networks, GPR requires less computational time and
	less input data to produce accurate results.
	
	\subsection{Gaussian Process Regression}
	
	Let $X=\left\{ \mathbf{x}_{i}\right\} _{i=1,\dots,n}\subset\mathbb{R}^{D}$
	be the set of predictors and $Y=\left\{ y_{i}\right\} _{i=1,\dots,n}\subset\mathbb{R}$
	the set of scalar outputs. These observations are modeled as the realization
	of the sum of a Gaussian process $\mathcal{G}$ and a Gaussian noise
	source $\varepsilon$. In particular, the distribution of $\mathbf{y}=\left(y_{1}\dots y_{n}\right)$
	is assumed to be given by 
	\begin{equation}
	\mathbf{y}\sim\mathcal{N}\left(\mu\left(X\right),K\left(X,X\right)+\sigma_{n}^{2}I_{n}\right),
	\end{equation}
	with $\mu$ the mean function, $I_{n}$ the $n\times n$ identity
	matrix and $K$ a $n\times n$ matrix given by $K\left(X,X\right)_{i,j}=k\left(\mathbf{x}_{i},\mathbf{x}_{j}\right)$.
	Here, we consider the Automatic Relevance Determination Squared Exponential
	(ARD SE) kernel, which is given by
	\begin{equation}
	k\left(\mathbf{x},\mathbf{x}'\right)=\sigma_{f}^{2}\exp\left(-\frac{1}{2}\sum_{k=1}^{D}\frac{1}{l_{k}^{2}}\left(x_{k}-x_{k}'\right)^{2}\right),
	\end{equation}
	where $\sigma_{f}^{2}$ is the signal variance and $l_{k}$ is the
	length-scale along the $k$ direction.
	
	Now, in addition, let us consider a test set $\tilde{X}$ of $m$
	points $\left\{ \tilde{\mathbf{x}}_{j}|j=1,\dots,m\right\} $. The
	realizations $\tilde{f}_{j}=\mathcal{G}\left(\tilde{\mathbf{x}}_{j}\right)+\varepsilon_{j}$
	are not known but are predicted through $\mathbb{E}\left[\mathbf{\tilde{f}}|\tilde{X},\mathbf{y},X\right]$:
	\begin{align}
	\mathbb{E}\left[\mathbf{\tilde{f}}|\tilde{X},\mathbf{y},X\right] & =\mu\left(\tilde{X}\right)+K\left(\tilde{X},X\right)A,\label{eq:GPR_prediction}
	\end{align}
	with $A=\left[K\left(X,X\right)+\sigma_{n}^{2}I_{n}\right]^{-1}\left(\mathbf{y}-\mu\left(X\right)\right)$.
	The mean function $\mu$, which is assumed to be a linear function
	of the predictors, is determined analytically by multi-linear regression.
	The parameters $\sigma_{f}^{2}$, $l_{1},\dots,l_{D}$ of the kernel
	and $\sigma_{n}^{2}$ of the noise are termed hyperparameters and
	they are estimated by log-likelihood maximization.
	
	The development of the GPR method consists in two steps: training
	and evaluation (also called testing). The former consists in estimating
	$\mu$, the hyperparameters and computing $A$ while the latter can
	be performed only after the training and it consists in obtaining
	the predictions via (\ref{eq:GPR_prediction}).
	
	\subsection{Applying the GPR to the GMWB Contract}
	
	We aim to apply the GPR method to GMWB products to speed up the computation
	of the price and of the Greeks. The modeling process starts by computing
	a training set $\mathcal{D}$. The predictor set $X$ consists of
	$n$ combinations of the stochastic model and GMWB product parameters.
	Specifically, the predictors are $v_{0}$, $k_{v}$, $\theta_{v}$,
	$\omega_{v}$, $\rho_{v}$, $r_{0}$, $k_{r}$, $\omega_{r}$, $\rho_{r}$
	$\alpha$ and $\kappa$. We accentuate that we can avoid considering
	the premium $P$ as a predictor since the price $\mathcal{V}$ of
	the GMWB contract is directly proportional to $P,$ that is $\mathcal{V}/P$
	does not depend on $P$. Similarly Greeks can be obtained from the
	Greeks computed for a particular value of $P$: for example, Delta
	does not depend on the considered value of $P$. Moreover we do not
	consider the contract maturity among the predictors since this is
	a discrete parameter which usually varies in a small set (usually
	$T=5,10$ or $20$) and one can compute an independent GPR for each
	value of $T$. Pseudo-random combinations are sampled uniformly over
	a fixed range for each parameter. Specifically, $n$ parameters combinations
	are sampled through the Faure sequence which covers efficiently all
	the domain of the parameters and yields better results than a random
	sample. For each parameter combination, we compute the price and then,
	observed data are passed to the GPR algorithm. Once the training step
	is finished, the model is ready to estimate prices. The same can be
	done for the Greeks. We point out that the procedure described above
	relates to the valuation of a single policy. If a portfolio of policies
	is considered instead, pricing and sensitivity calculation can be
	done for each contract individually. The no-arbitrage fee $\alpha_{na}$
	of a GMWB contract is the particular value of $\alpha$ which makes
	the initial value of the policy $\mathcal{V}\left(P,P,v_{0},r_{0},0\right)$
	equal to the premium $P$ and its computation is a common practice
	before the sale of a contract. This value is usually determined by
	employing the secant method, seeking to equate $\mathcal{V}\left(P,P,v_{0},r_{0},0\right)$
	and $P$. The GPR method can be applied within the secant method to
	boost the computation of $\alpha_{na}$ by replacing the direct computation
	of $\mathcal{V}\left(P,P,v_{0},r_{0},0\right)$ with HPDE by the GPR
	prediction.
	
	\FloatBarrier
	
	\section{Numerical Results}
	
	In this Section we report some numerical results. Specifically, in
	the first test we show the accuracy of the Hybrid Tree-PDE algorithm
	in pricing and in computing Delta and the no-arbitrage fees. Then,
	in the following two tests, we apply the GPR method to predict the
	no-arbitrage fee and the Greek $\Delta$ of a GMWB contract respectively.
	The HPDE algorithm has been implemented in C++, whereas the regression
	algorithm has been implemented in MATLAB and computations have been
	performed on a PC equipped with $8$ GB of RAM and a 2.5GHz i5-7200u
	processor. The performance of the GPR mehtod is measured in terms
	of the following indicators which depend on the maximum and average
	absolute and relative error: the RMSE (Root Mean Squared Error), the
	RMSRE (Root Mean Squared Relative Error), the MaxAE (Maximum Absolute
	Error) and the MaxRE (Maximum Relative Error). Finally, we also report
	the speed-up that is the ratio between the computational time achieved
	with a direct computation by HPDE and the computational time achieved
	by using the GPR method.
	
	\subsection{Testing the HPDE method }
	
	We show the accuracy of the HPDE method which is involved in the computation
	of the data employed in both the training and testing steps. Input
	parameters are reported in Table \ref{tab:Par_HHW} while results
	are available in Table \ref{tab:Test0}. Moreover, the DAV 2004R mortality
	Table for a 65 year old German male is employed to compute the functions
	$\mathcal{M}$ and $\mathcal{R}$ (see Forsyth and Vetzal \cite{forsyth2014optimal}
	for the Table). Specifically, we compute the price, the Delta and
	the no-arbitrage fee considering several configurations with increasing
	number of time and space steps (see Gouden\`ege et al. \cite{goudenege2016,goudenege2019}
	for further details). Numerical results suggested that a few steps
	is enough to obtain accurate values but, the computational times may
	become considerable. 
	
	\begin{table}
		\begin{centering}
			{\small{}}%
			\begin{tabular}{lccclcc}
				{\small{}Name} & {\small{}Symbol} & {\small{}Value} &  & {\small{}Name} & {\small{}Symbol} & {\small{}Value}\tabularnewline
				\midrule
				{\small{}Premium} & {\small{}$P$} & {\small{}$100$} &  & {\small{}Initial interest rate} & {\small{}$r_{0}$} & {\small{}$0.02$}\tabularnewline
				{\small{}Maturity} & {\small{}$T$} & {\small{}$10$ years} &  & {\small{}Rate of mean reversion } & {\small{}$k_{r}$} & {\small{}$0.15$}\tabularnewline
				{\small{}Initial volatility} & {\small{}$v_{0}$} & {\small{}$0.05$} &  & {\small{}Volatility of interest rate} & {\small{}$\omega_{r}$} & {\small{}$0.015$}\tabularnewline
				{\small{}Rate of mean reversion} & {\small{}$k_{v}$} & {\small{}$2.00$} &  & {\small{}Correlation} & {\small{}$\rho_{r}$} & {\small{}$0.20$}\tabularnewline
				{\small{}Long run variance} & {\small{}$\theta_{v}$} & {\small{}$0.05$} &  & {\small{}Fees} & {\small{}$\alpha$} & {\small{}$0.035$}\tabularnewline
				{\small{}Volatility of volatility} & {\small{}$\omega_{v}$} & {\small{}$0.50$} &  & {\small{}Penalty} & {\small{}$\kappa$} & {\small{}$0.10$}\tabularnewline
				{\small{}Correlation} & {\small{}$\rho_{v}$} & {\small{}$-0.55\phantom{-}$} &  &  &  & \tabularnewline
				\bottomrule
			\end{tabular}
			\par\end{centering}{\small \par}
		\caption{\label{tab:Par_HHW}The parameters employed.}
	\end{table}
	
	\begin{table}
		\begin{centering}
			{\small{}}%
			\begin{tabular}{lcccc}
				{\small{}Time and space steps} & {\small{}$125\times125$} & {\small{}$250\times250$} & {\small{}$500\times500$} & {\small{}$1000\times1000$}\tabularnewline
				\midrule 
				{\small{}Price} & {\small{}$\underset{\left(7.8\text{e}+0\right)}{100.08}$ } & {\small{}$\underset{\left(8.3\text{e}+1\right)}{100.11}$ } & {\small{}$\underset{\left(1.1\text{e}+3\right)}{100.12}$ } & {\small{}$\underset{\left(1.4\text{e}+4\right)}{100.12}$ }\tabularnewline
				{\small{}Delta} & {\small{}$\underset{\left(7.8\text{e}+0\right)}{0.3881}$ } & {\small{}$\underset{\left(8.3\text{e}+1\right)}{0.3888}$ } & {\small{}$\underset{\left(1.1\text{e}+3\right)}{0.3892}$ } & {\small{}$\underset{\left(1.4\text{e}+4\right)}{0.3894}$ }\tabularnewline
				{\small{}No-arbitrage fee } & {\small{}$\underset{\left(4.6\text{e}+1\right)}{353.45}$ } & {\small{}$\underset{\left(5.0\text{e}+2\right)}{354.81}$ } & {\small{}$\underset{\left(6.6\text{e}+3\right)}{355.25}$} & {\small{}$\underset{\left(8.4\text{e}+4\right)}{355.55}$}\tabularnewline
				\bottomrule
			\end{tabular}
			\par\end{centering}{\small \par}
		\caption{\label{tab:Test0}Hybrid Tree-PDE computation results. The values
			in brackets are the computational times measured in seconds. The no-arbitrage
			fee is expressed in bps.}
		
	\end{table}
	
	\subsection{No-arbitrage Fee Results}
	
	We test the ability of the GPR model to compute the no-arbitrage fee
	of a GMWB contract. Table \ref{tab:RangeHHW} report the ranges for
	the input parameters. The training set consists of $n=1250,$ $2500$,
	$5000$, $10000$, or $20000$ combinations of parameters and the
	prices obtained by using the HPDE method with $250$ time and space
	steps.  The regression model is tested on both input data and out-of-sample
	data. In particular, out-of-sample data consists of $m=20000$ additional
	parameters combinations which are obtained by a random simulation.
	Numerical results are reported in Table \ref{tab:Results_fee}. The
	scatter plot of out-of-sample absolute errors for a model trained
	on $n=10000$ data is shown in Figure \ref{fig:Figure-3}. Specifically,
	prediction errors are sorted according to the respective policy price
	and we can see that the errors remain in an admissible band.
	
	\begin{table}
		\begin{centering}
			\setlength\tabcolsep{4pt}{\small{}}%
			\begin{tabular}{lccclcc}
				{\small{}Name} & {\small{}Symbol} & {\small{}Value} &  & {\small{}Name} & {\small{}Symbol} & {\small{}Value}\tabularnewline
				\midrule
				{\small{}Premium} & {\small{}$P$} & {\small{}$100$} &  & {\small{}Initial interest rate} & {\small{}$r_{0}$} & {\small{}$\left[0.01,0.03\right]$}\tabularnewline
				{\small{}Maturity} & {\small{}$T$} & {\small{}$10$ years} &  & {\small{}Rate of mean reversion } & {\small{}$k_{r}$} & {\small{}$\left[0.05,0.25\right]$}\tabularnewline
				{\small{}Initial volatility} & {\small{}$v_{0}$} & {\small{}$\left[0.01,0.10\right]$} &  & {\small{}Volatility of interest rate} & {\small{}$\omega_{r}$} & {\small{}$\left[0.005,0.025\right]$}\tabularnewline
				{\small{}Rate of mean reversion} & {\small{}$k_{v}$} & {\small{}$\left[1.40,2.60\right]$} &  & {\small{}Correlation} & {\small{}$\rho_{r}$} & {\small{}$\left[0.05,0.35\right]$}\tabularnewline
				{\small{}Long run variance} & {\small{}$\theta_{v}$} & {\small{}$\left[0.01,0.10\right]$} &  & {\small{}Fees} & {\small{}$\alpha$} & {\small{}$\left[0.00,0.10\right]$}\tabularnewline
				{\small{}Volatility of volatility} & {\small{}$\omega_{v}$} & {\small{}$\left[0.45,0.75\right]$} &  & {\small{}Penalty} & {\small{}$\kappa$} & {\small{}$\left[0.00,0.20\right]$}\tabularnewline
				{\small{}Correlation} & {\small{}$\rho_{v}$} & {\small{}$\left[\lyxmathsym{­}-0.70,-0.40\right]$} &  &  &  & \tabularnewline
				\bottomrule
			\end{tabular}
			\par\end{centering}{\small \par}
		\caption{\label{tab:RangeHHW}The range of the parameters.}
	\end{table}
	
	\begin{table}
		\begin{centering}
			{\small{}}%
			\begin{tabular}{clllll}
				{\small{}Size of training set} & {\small{}1250} & {\small{}2500} & {\small{}5000} & {\small{}10000} & {\small{}20000}\tabularnewline
				\midrule
				{\small{}RMSE} & {\small{}7.88e-4} & {\small{}5.94e-4} & {\small{}4.82e-4} & {\small{}3.93e-4} & {\small{}3.74e-4}\tabularnewline
				{\small{}RMSRE} & {\small{}2.30e-2} & {\small{}1.98e-2} & {\small{}1.54e-2} & {\small{}1.36e-2} & {\small{}1.20e-2}\tabularnewline
				{\small{}MaxAE} & {\small{}2.02e-2} & {\small{}1.08e-2} & {\small{}8.10e-3} & {\small{}6.50e-3} & {\small{}6.13e-3}\tabularnewline
				{\small{}MaxRE} & {\small{}3.22e-1} & {\small{}3.07e-1} & {\small{}2.74e-1} & {\small{}1.66e-1} & {\small{}1.51e-1}\tabularnewline
				\midrule
				{\small{}Speed-up} & {\small{}$\times$9.8e5} & {\small{}$\times$3.8e5} & {\small{}$\times$1.8e5} & {\small{}$\times$1.2e5} & {\small{}$\times$6.4e4}\tabularnewline
				\bottomrule
			\end{tabular}
			\par\end{centering}{\small \par}
		\caption{\label{tab:Results_fee}Performances of the GPR method in computing
			the no-arbitrage fee.}
	\end{table}
	
	\subsection{Delta Results}
	
\begin{figure}
		\begin{centering}
			\subfloat[\label{fig:Figure-3}No-arbitrage fee ]{\begin{centering}
					\includegraphics[width=0.45\textwidth]{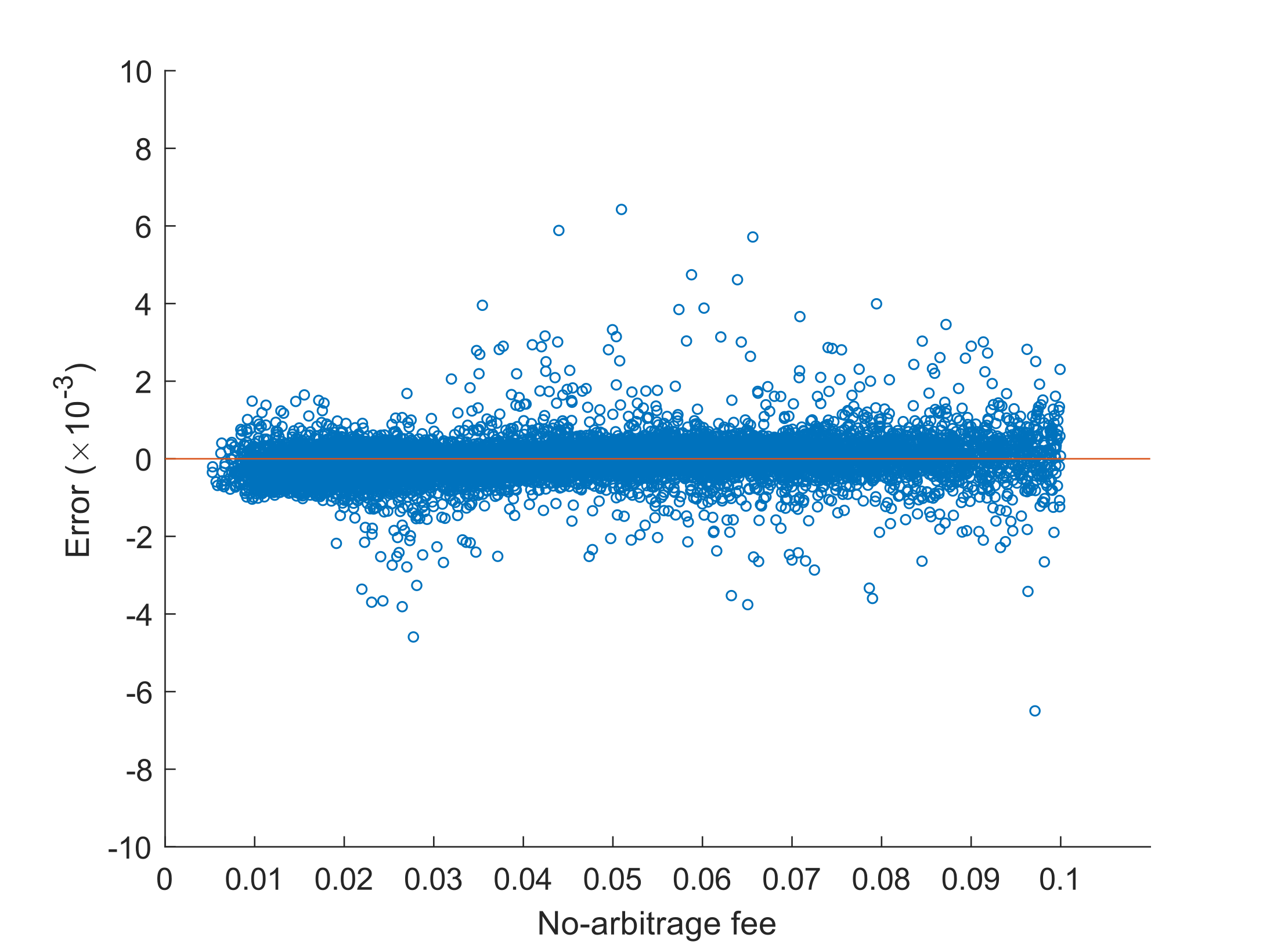}
					\par\end{centering}
			}\subfloat[\label{fig:Figure-2} Delta]{\begin{centering}
					\includegraphics[width=0.45\textwidth]{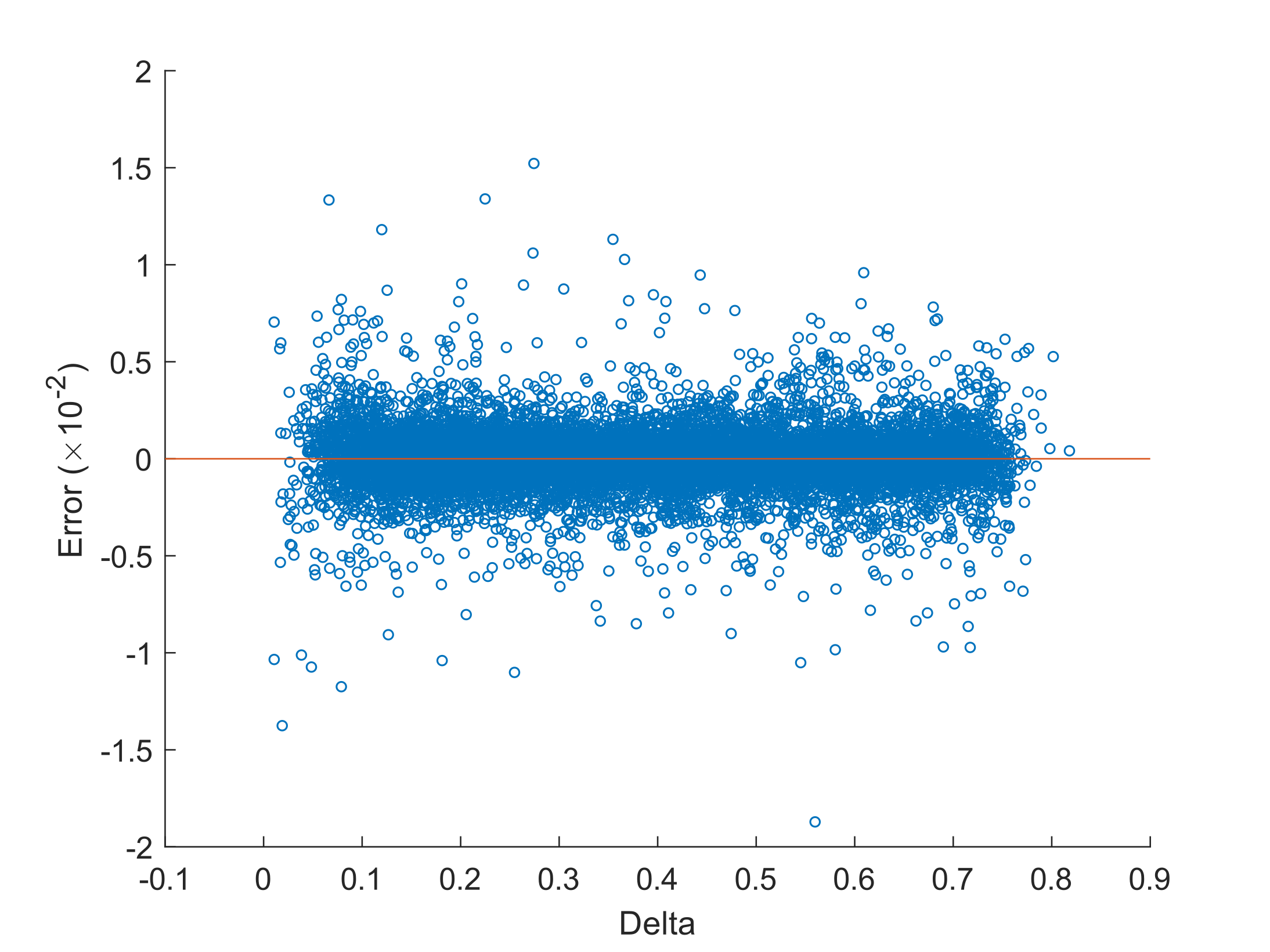}
					\par\end{centering}
				
			}
			\par\end{centering}
		\caption{Out-of-sample prediction absolute errors for a GPR method trained
			on $n=10000$ points. }
	\end{figure}

	We test the ability of the GPR model to predict the first derivative
	Delta of a GMWB contract, which is crucial in hedging. Numerical results
	are reported in Table \ref{tab:Results_Delta}, while the scatter
	plot of the estimation errors is shown in Figure \ref{fig:Figure-2}.
	In particular, errors are distributed around zero with any evident
	outline.
	
	\begin{table}
		\begin{centering}
			{\small{}}%
			\begin{tabular}{lclllll}
				\multicolumn{2}{l}{{\small{}Size of training set}} & {\small{}1250} & {\small{}2500} & {\small{}5000} & {\small{}10000} & {\small{}20000}\tabularnewline
				\midrule
				& {\small{}RMSE} & {\small{}1.94e-4} & {\small{}6.09e-4} & {\small{}6.93e-4} & {\small{}8.16e-4} & {\small{}9.37e-4}\tabularnewline
				{\small{}In-sample } & {\small{}RMSRE} & {\small{}8.28e-4} & {\small{}3.75e-3} & {\small{}3.95e-3} & {\small{}5.16e-3} & {\small{}6.07e-3}\tabularnewline
				{\small{}prediction} & {\small{}MaxAE} & {\small{}8.84e-4} & {\small{}3.95e-3} & {\small{}6.30e-3} & {\small{}7.78e-3} & {\small{}1.60e-2}\tabularnewline
				& {\small{}MaxRE} & {\small{}7.47e-3} & {\small{}9.42e-2} & {\small{}1.08e-1} & {\small{}1.64e-1} & {\small{}2.52e-1}\tabularnewline
				\midrule
				& {\small{}RMSE} & {\small{}2.65e-3} & {\small{}1.88e-3} & {\small{}1.44e-3} & {\small{}1.22e-3} & {\small{}1.16e-3}\tabularnewline
				{\small{}Out-of-sample } & {\small{}RMSRE} & {\small{}1.83e-2} & {\small{}1.20e-2} & {\small{}8.97e-3} & {\small{}7.78e-3} & {\small{}7.39e-3}\tabularnewline
				{\small{}prediction} & {\small{}MaxAE} & {\small{}3.26e-2} & {\small{}2.70e-2} & {\small{}2.47e-2} & {\small{}1.87e-2} & {\small{}2.10e-2}\tabularnewline
				& {\small{}MaxRE} & {\small{}6.25e-1} & {\small{}4.26e-1} & {\small{}2.64e-1} & {\small{}2.63e-1} & {\small{}2.50e-1}\tabularnewline
				\midrule
				\multicolumn{2}{l}{{\small{}Speed-up}} & {\small{}$\times$9.8e5} & {\small{}$\times$3.8e5} & {\small{}$\times$1.8e5} & {\small{}$\times$1.2e5} & {\small{}$\times$6.4e4}\tabularnewline
				\bottomrule
			\end{tabular}
			\par\end{centering}{\small \par}
		\caption{\label{tab:Results_Delta} Performances of the GPR method in computing
			Delta.}
	\end{table}
	
	\subsection{Remarks on Computational Results}
	
	Tables \ref{tab:Results_fee} and \ref{tab:Results_Delta} show the
	performance of the developed approach. Errors are very small in all
	the considered cases, indicating that the predictions are very accurate.
	The most interesting aspect is the gain in terms of computation time:
	it is reduced by thousands of times. Moreover, Table \ref{tab:Training_com_cost}
	shows the computational time for the training step. The cost of the
	training step depends on both the size of the training set, and the
	target function (price or Delta). We observe that the required time
	is high when the size of the training set is $n=20000$ because in
	this case the BCD algorithm is used in place of the exact computation
	(see Grippo and Sciandrone \cite{grippo2000}). Finally, Table \ref{tab:Testing_results}
	shows the predicted price, Delta and no-arbitrage fee for the same
	GMWB policy considered in Table \ref{tab:Test0}. We observe that
	the estimates are very accurate and the gain in terms of computational
	time is considerable.
	
	\begin{table}
		\begin{centering}
			{\small{}}%
			\begin{tabular}{cccccc}
				{\small{}Size of training set} & {\small{}1250} & {\small{}2500} & {\small{}5000} & {\small{}10000} & {\small{}20000}\tabularnewline
				\midrule 
				{\small{}Price} & {\small{}49} & {\small{}154} & {\small{}185} & {\small{}188} & {\small{}3219}\tabularnewline
				{\small{}Delta} & {\small{}32} & {\small{}110} & {\small{}128} & {\small{}142} & {\small{}3144}\tabularnewline
				\bottomrule
			\end{tabular}
			\par\end{centering}{\small \par}
		\caption{\label{tab:Training_com_cost}The computational time (in seconds)
			to train the GPR algorithm.}
	\end{table}
	
	\begin{table}
		\begin{centering}
			{\small{}}%
			\begin{tabular}{cccccc}
				{\small{}Size of training set} & {\small{}1250} & {\small{}2500} & {\small{}5000} & {\small{}10000} & {\small{}20000}\tabularnewline
				\midrule 
				{\small{}Price} & {\small{}$\underset{\left(8.5\text{e}-5\right)}{100.10}$ } & {\small{}$\underset{\left(2.2\text{e}-4\right)}{100.09}$ } & {\small{}$\underset{\left(4.6\text{e}-4\right)}{100.09}$ } & {\small{}$\underset{\left(6.8\text{e}-4\right)}{100.09}$ } & {\small{}$\underset{\left(1.3\text{e}-3\right)}{100.09}$ }\tabularnewline
				{\small{}Delta} & {\small{}$\underset{\left(8.4\text{e}-5\right)}{0.3899}$ } & {\small{}$\underset{\left(1.9\text{e}-4\right)}{0.3892}$ } & {\small{}$\underset{\left(4.2\text{e}-4\right)}{0.3892}$ } & {\small{}$\underset{\left(7.4\text{e}-4\right)}{0.3891}$ } & {\small{}$\underset{\left(1.4\text{e}-3\right)}{0.3891}$ }\tabularnewline
				{\small{}No-arbitrage fee} & {\small{}$\underset{\left(5.1\text{e}-4\right)}{354.27}$ } & {\small{}$\underset{\left(1.3\text{e}-3\right)}{354.08}$ } & {\small{}$\underset{\left(2.8\text{e}-3\right)}{353.93}$ } & {\small{}$\underset{\left(4.1\text{e}-3\right)}{354.01}$ } & {\small{}$\underset{\left(7.8\text{e}-3\right)}{354.03}$ }\tabularnewline
				\bottomrule
			\end{tabular}
			\par\end{centering}{\small \par}
		\caption{\label{tab:Testing_results}GPR computation results. The values in
			brackets are the computational times measured in seconds. The no-arbitrage
			fee is expressed in bps.}
	\end{table}
	
	\section{Conclusions}
	
	In this paper we have presented how Gaussian Process Regression can
	be applied in the insurance field to address the problem of pricing
	and computing the sensitivities of a GMWB Variable Annuity when both
	stochastic volatility and stochastic interest rate are considered.
	The HPDE method proves to be an effective tool for calculating prices
	and Greeks in the Heston Hull-White model, however it may be time
	demanding if a large number of time-steps is considered. The GPR method
	allows one to considerably reduce the computational time as the greater
	computational effort is carried out during the training phase, which
	must be performed only once before the actual use of the model while
	computing a single prediction. The use of a pseudo-random method and
	of a general kernel allow one to obtain accurate results with a small
	amount of observations. The resulting model can be applied to compute
	the prices of the policies, for hedging purposes, or to compute the
	no-arbitrage cost of a policy and can be particularly useful in the
	case of books of policies, when the evaluation must be repeated many
	times. We conclude observing that the same approach may be applied
	to all types of Variable Annuities contracts. 
	
	\appendix
	
	\section{\label{sec:Ap} The Trinomial Tree }
	
	In this Appendix we present how to build a trinomial tree for the
	volatility process $v$ and the interest rate process $x$. In the
	scope of pricing long maturity products, the trinomial tree proposed
	here turns out to be more suitable of the binomial tree proposed by
	Briani et al. \cite{briani2018} since it matches exactly the first
	two moments of the developed process and convergence is faster. Let
	$Z$ be a Brownian motion and let $G$ be a Gaussian process, given
	by 
	\begin{equation}
	dG_{t}=a\left(G_{t}\right)dt+bdZ_{t},
	\end{equation}
	with variance that depends only by the time lapse, i.e. $G_{t+h}|\mathcal{F}_{t}\sim\mathcal{N}\left(\mu\left(h,G_{t}\right),\sigma^{2}\left(h\right)\right)$
	where $\mu\left(h,G_{t}\right)$ and $\sigma^{2}\left(h\right)$ are
	respectively the expectation and the variance of the of the process
	$G$.
	
	We show how to build a simple trinomial tree that can match the first
	two moments of $G$. Let's fix a maturity $T$ a number of time-steps
	$N$ and setfine $\Delta t=\nicefrac{T}{N}$. Each node will be denoted
	by $G_{n,j}$ where $n\in\left\{ 0,\dots,N\right\} $ and $j\in\left\{ 0,\dots,2n\right\} $
	. The value of each node is 
	\begin{equation}
	G_{n,j}=G_{0}+\frac{3}{2}\left(j-n\right)\sqrt{\sigma^{2}\left(\Delta t\right)}.
	\end{equation}
	We recall the first two moments $\left(M_{1},M_{2}\right)$ of the
	process $G$: 
	\begin{equation}
	M_{1}\left(\Delta t,G_{t}\right)=\mathbb{E}\left[G_{t+\Delta t}|\mathcal{F}_{t}\right]=\mu\left(\Delta t,G_{t}\right),\ M_{2}\left(\Delta t,G_{t}\right)=\mathbb{E}\left[\left(G_{t+\Delta t}\right)^{2}|\mathcal{F}_{t}\right]=\mu^{2}\left(\Delta t,G_{t}\right)+\sigma^{2}\left(\Delta t\right).
	\end{equation}
	
	Let's fix a node $G_{n,j}$. To be brief, $\mu$ will denote $\mu\left(\Delta t,G_{n,j}\right)$
	and $\sigma$ will denote $\sqrt{\sigma^{2}\left(\Delta t\right)}$.
	We suppose that the conditional expected value $\mu=\mathbb{E}\left[G_{t+h}|\mathcal{F}_{t}\right]$
	falls between the values of the nodes at time $\left(n+1\right)\Delta t$.
	This hypothesis can be justify assuming that the time increment $\Delta t$
	is small enough (in fact, by continuity, $\lim_{\Delta t\rightarrow0^{+}}\mu\left(\Delta t,G_{n,j}\right)=G_{n,j}$
	which is a node). We define 
	\begin{equation}
	j_{A}\left(n,j\right)=n+\mbox{ceil}\left[\frac{2}{3\sigma}\left(\mu-G_{0}\right)\right],
	\end{equation}
	i.e. the first node whose value is bigger than the mean of the process.
	Let 
	\begin{equation}
	j_{B}\left(n,j\right)=j_{A}\left(n,j\right)-1,\ j_{C}\left(n,j\right)=j_{A}\left(n,j\right)+1,\ j_{D}\left(n,j\right)=j_{A}\left(n,j\right)-2.
	\end{equation}
	To be brief we will only write $j_{A},j_{B},j_{C},$ $j_{D}$, and
	$G_{A}$ will be $G_{A}=G_{n+1,j_{A}}$ and the same for the other
	letters. We can now define a Markovian discrete time process $\hat{G}_{n}$,
	$n=0,\dots,N$ with $\hat{G}_{0}=G_{0,0}$. Let $\hat{G}_{n}=G_{n,j}$.
	If $0\leq G_{A}-\mu\leq\frac{3}{4}\sigma$ then $\hat{G}_{n}$ can
	move to $G_{A}$, $G_{B}$, $G_{C}$, otherwise $\hat{G}_{n}$ can
	move to $G_{A}$, $G_{B}$, $G_{D}$. Transition probabilities are
	stated in Table \ref{tab:Transition-probabilities-for}.
	
	\begin{table}
		\begin{centering}
			\begin{tabular}{ccc}
				& if $0\leq G_{A}-\mu\leq\frac{3}{4}\sigma$ & if $\frac{3}{4}\sigma<G_{A}-\mu$\tabularnewline
				\midrule
				$p_{A}$ & $\frac{5\sigma^{2}-4\left(G_{A}-\mu\right)^{2}}{9\sigma^{2}}$ & $\frac{2\left(\mu-G_{B}\right)^{2}+3\sigma\left(\mu-G_{B}\right)+2\sigma^{2}}{9\sigma^{2}}$\tabularnewline
				$p_{B}$ & $\frac{2\left(G_{A}-\mu\right)^{2}+3\sigma\left(G_{A}-\mu\right)+2\sigma^{2}}{9\sigma^{2}}$ & $\frac{5\sigma^{2}-4\left(\mu-G_{B}\right)^{2}}{9\sigma^{2}}$\tabularnewline
				$p_{C}$ & $\frac{2\left(G_{A}-\mu\right)^{2}-3\sigma\left(G_{A}-\mu\right)+2\sigma^{2}}{9\sigma^{2}}$ & $0$\tabularnewline
				$p_{D}$ & $0$ & $\frac{2\left(\mu-G_{B}\right)^{2}-3\sigma\left(\mu-G_{B}\right)+2\sigma^{2}}{9\sigma^{2}}$\tabularnewline
				\bottomrule
			\end{tabular}
			\par\end{centering}
		\caption{\label{tab:Transition-probabilities-for}Transition probabilities
			for the trinomial tree.}
		
	\end{table}
	Since $G_{B}<\mu\leq G_{A}$, we can easily show that these probabilities
	are well defined: all in $\left[0,1\right]$, their sum is equal to
	$1$, and the first two moments of the variable $\hat{G}_{n+1}|\hat{G}_{n}=G_{n,j}$
	are equal to the first two moments of the variable $G_{t+h}|G_{t}=G_{n,j}$.
	
	We can approximate the process $G$ by a process $\bar{G}$ that is
	constant in each time lapse and is defined by $\bar{G}_{t}=\hat{G}_{\left\lfloor \nicefrac{t}{h}\right\rfloor }$.
	The weak convergence of this tree can be proved as in Nelson and Ramaswamy
	\cite{nelson1990simple}.
	
	This construction can be directly applied to build a tree for the
	process $x$ since it is Gaussian. As far as the volatility process
	is concerned, this method cannot be directly applied since $v$ has
	no constant variance and is not Gaussian. However, it can be easily
	be modified as in \cite{goudenege2016}, so as to guarantee the week
	convergence.
	
	\bibliographystyle{abbrv}
	\bibliography{bibliography}
	
\end{document}